\documentclass[10pt,conference]{IEEEtran}

\usepackage{cite}
\usepackage{amsmath}
\usepackage{graphicx}
\usepackage{booktabs}
\usepackage{array}
\usepackage{multirow}
\usepackage{xcolor}
\usepackage{xurl}
\usepackage[hidelinks]{hyperref}

\newcommand{\NPackagist}{460,841}
\newcommand{\NRetrieved}{460,792}
\newcommand{\NGone}{49}
\newcommand{\ListedOn}{6 September 2026}
\newcommand{\NLegacy}{202,136}
\newcommand{\NIndexed}{186,306}
\newcommand{\NComposerJson}{185,817}
\newcommand{\NLockless}{154,004}
\newcommand{\NTests}{60,202}
\newcommand{\NFrame}{4,868}
\newcommand{\LocklessShare}{82.9\%}
\newcommand{\FrameSeed}{20261001}
\newcommand{\ComposerTwo}{2.10.3}
\newcommand{\ComposerOne}{1.10.28}
\newcommand{\PHPSeven}{7.4.33}

\newcommand{\NProcessed}{4,868}
\newcommand{\NWithOutcomes}{4,698}
\newcommand{\NDeclaresRepos}{170}

\newcommand{\PrimN}{3,660}
\newcommand{\PrimRecovered}{349}
\newcommand{\PrimRegressed}{88}
\newcommand{\PrimDiff}{7.1\,pp}
\newcommand{\PrimDiffLo}{6.0}
\newcommand{\PrimDiffHi}{8.2}
\newcommand{\PrimP}{$p<0.001$}
\newcommand{\PracN}{3,778}

\newcommand{\PracDiff}{10.9\,pp}

\newcommand{\ToolRecovered}{157}

\newcommand{\ToolDiff}{3.9\,pp}

\newcommand{\MetaOneN}{3,766}

\newcommand{\MetaOneDiff}{9.4\,pp}

\newcommand{\ToolModernDiff}{1.3\,pp}

\newcommand{\OrigToolDiff}{$-$7.1\,pp}

\newcommand{\FalseBaseNum}{361}
\newcommand{\FalseBaseDen}{3,659}
\newcommand{\FalseBaseShare}{9.9\%}
\newcommand{\FalseBaseLo}{8.9\%}
\newcommand{\FalseBaseHi}{10.9\%}
\newcommand{\DriftCompared}{3,530}

\newcommand{\DriftDifferShare}{98.5\%}

\newcommand{\DriftMajorShare}{80.2\%}
\newcommand{\DriftChangedMedian}{53}
\newcommand{\FlakyRerun}{3,030}
\newcommand{\FlakyN}{38}
\newcommand{\SensExecN}{2,873}
\newcommand{\SensExecRecovered}{349}
\newcommand{\SensExecRegressed}{88}
\newcommand{\SensExecDiff}{9.1\,pp}
\newcommand{\SensExecDiffLo}{7.7}
\newcommand{\SensExecDiffHi}{10.5}
\newcommand{\SensStableN}{3,629}
\newcommand{\SensStableRecovered}{346}
\newcommand{\SensStableRegressed}{86}
\newcommand{\SensStableDiff}{7.2\,pp}
\newcommand{\SensStableDiffLo}{6.1}
\newcommand{\SensStableDiffHi}{8.3}
\newcommand{\SensSkipN}{3,654}
\newcommand{\SensSkipRecovered}{349}
\newcommand{\SensSkipRegressed}{87}
\newcommand{\SensSkipDiff}{7.2\,pp}
\newcommand{\SensSkipDiffLo}{6.1}
\newcommand{\SensSkipDiffHi}{8.3}
\newcommand{\SensCommitN}{3,152}
\newcommand{\SensCommitRecovered}{288}
\newcommand{\SensCommitRegressed}{84}
\newcommand{\SensCommitDiff}{6.5\,pp}
\newcommand{\SensCommitDiffLo}{5.3}
\newcommand{\SensCommitDiffHi}{7.6}

\newcommand{\DescBoneGainsPlugin}{148}
\newcommand{\DescBtwoLossesPlugin}{51}

\newcommand{\DescBtwoLossesRescued}{50}

\newcommand{\Recoveries}{349}
\newcommand{\RecovFail}{279}
\newcommand{\RecovFatal}{61}
\newcommand{\RecovNoReport}{9}
\newcommand{\AttrAttributed}{334}

\newcommand{\AttrCoMoved}{230}

\newcommand{\ConfPartial}{8}

\newcommand{\ConfCases}{342}

\newcommand{\ConfSignature}{301}

\newcommand{\ManualN}{20}

\newcommand{\ManualKappaCulprit}{0.79}

\newcommand{\ManualImplicated}{16}

\newcommand{\EtwoManualN}{20}

\newcommand{\ValRqOneN}{20}
\newcommand{\ValRqThreeN}{40}

\newcommand{\ValCulpritYes}{19}

\newcommand{\ValKappaCulprit}{0.31}
\newcommand{\ValKappaCulpritLo}{$-$0.05}
\newcommand{\ValKappaCulpritHi}{0.48}
\newcommand{\ValKappaLoc}{1.00}

\newcommand{\ValKappaDiff}{0.27}
\newcommand{\ValKappaDiffLo}{0.14}
\newcommand{\ValKappaDiffHi}{0.36}

\newcommand{\ProbeN}{12}

\newcommand{\ProbeAgentMissedMax}{6}

\newcommand{\EoneWithLock}{31,813}
\newcommand{\EoneLockTests}{20,079}
\newcommand{\EoneCandidates}{1,204}
\newcommand{\EoneExcludedRepos}{59}
\newcommand{\EoneUnparseable}{1}

\newcommand{\EoneFrame}{1,144}

\newcommand{\EoneGraphTone}{616}
\newcommand{\EoneInfraBtwo}{27}
\newcommand{\EoneInfraAtwo}{28}
\newcommand{\EoneTone}{617}
\newcommand{\EoneToneN}{436}

\newcommand{\EoneTtwo}{898}
\newcommand{\EoneTtwoN}{616}

\newcommand{\EoneTtwoDelta}{64.7\,pp}
\newcommand{\EoneTthree}{1,144}
\newcommand{\EoneTthreeN}{812}

\newcommand{\EoneTthreeDelta}{59.9\,pp}
\newcommand{\EoneHN}{436}
\newcommand{\EoneHCloser}{427}
\newcommand{\EoneHFurther}{0}
\newcommand{\EoneHTies}{9}
\newcommand{\EoneHP}{$p<0.001$}
\newcommand{\EoneDelta}{67.1\,pp}
\newcommand{\EoneDeltaLo}{64.4}
\newcommand{\EoneDeltaHi}{69.6}

\newcommand{\EoneSensN}{588}
\newcommand{\EoneSensCloser}{427}
\newcommand{\EoneSensFurther}{18}
\newcommand{\EoneSensDelta}{49.4\,pp}

\newcommand{\EoneMBtwo}{80.7\%}
\newcommand{\EoneExactBtwo}{57}
\newcommand{\EoneLockBtwo}{436}
\newcommand{\EoneUnresBtwo}{153}

\newcommand{\EoneMAtwo}{13.5\%}
\newcommand{\EoneExactAtwo}{2}
\newcommand{\EoneLockAtwo}{455}
\newcommand{\EoneUnresAtwo}{133}

\newcommand{\EoneMP}{22.6\%}
\newcommand{\EoneExactP}{2}
\newcommand{\EoneLockP}{455}
\newcommand{\EoneUnresP}{133}

\newcommand{\EoneMBL}{88.9\%}
\newcommand{\EoneExactBL}{114}
\newcommand{\EoneLockBL}{461}
\newcommand{\EoneUnresBL}{155}

\newcommand{\EoneSoneN}{286}
\newcommand{\EoneSoneBtwo}{36}

\newcommand{\EoneSoneP}{$p<0.001$}

\newcommand{\EoneFBAtwoShare}{24.5\%}

\newcommand{\EoneFBBtwo}{44}
\newcommand{\EoneFBBtwoDen}{326}
\newcommand{\EoneFBBtwoShare}{13.5\%}

\newcommand{\EoneFBPShare}{35.0\%}

\newcommand{\EoneNewerBtwo}{3,494}
\newcommand{\EoneOlderBtwo}{20}

\newcommand{\EoneBLCloser}{218}
\newcommand{\EoneBLFurther}{2}

\newcommand{\EoneAbsentProjects}{61}

\newcommand{\EoneLLocks}{580}

\newcommand{\EoneLibShare}{84\%}
\newcommand{\EoneYearShare}{96\%}

\newcommand{\MainLibShare}{79\%}
\newcommand{\MainYearShare}{97\%}

\newcommand{\EoneCommitDelta}{67.0\,pp}

\newcommand{\EoneFBBtwoUnresolvable}{41}

\newcommand{\EtwoPairs}{1,638}
\newcommand{\EtwoAEligible}{1,191}
\newcommand{\EtwoANoTask}{760}
\newcommand{\EtwoAMasked}{178}
\newcommand{\EtwoAManufactured}{1}
\newcommand{\EtwoATransformed}{125}
\newcommand{\EtwoAStable}{127}

\newcommand{\EtwoADall}{25.5\%}

\newcommand{\EtwoADexist}{15.0\%}
\newcommand{\EtwoADexistN}{179}
\newcommand{\EtwoADexistLo}{13.1\%}
\newcommand{\EtwoADexistHi}{17.2\%}

\newcommand{\EtwoAP}{$p<0.001$}

\newcommand{\EtwoABothFail}{252}

\newcommand{\EtwoAFailingBtwo}{430}

\newcommand{\EtwoAFailingAtwo}{253}

\newcommand{\EtwoBDexist}{9.2\%}

\newcommand{\EtwoMovedBtwo}{1,083}
\newcommand{\EtwoMovedAtwo}{221}
\newcommand{\EtwoPreserving}{56}

\newcommand{\EtwoPreservingMasked}{3}

\newcommand{\EtwoSfiveBtwoAll}{1,191}

\newcommand{\EtwoSfiveBoth}{738}

\newcommand{\EtwoRtwoBoth}{740}
\newcommand{\EtwoRtwoSame}{737}

\newcommand{\EtwoARepeated}{304}

\newcommand{\EtwoMechMaskedDepr}{130}
\newcommand{\EtwoMechLaravel}{129}
\newcommand{\EtwoMechHttpFoundation}{121}

\newcommand{\EtwoMechRestN}{1,061}

\newcommand{\EtwoMechRestDexist}{4.6\%}
\newcommand{\EtwoMechRestDexistLo}{3.5\%}
\newcommand{\EtwoMechRestDexistHi}{6.1\%}

\newcommand{\EtwoFwLaravelN}{287}
\newcommand{\EtwoFwLaravelChange}{131}
\newcommand{\EtwoFwLaravelRate}{45.6\%}

\newcommand{\EtwoFwNoneN}{699}
\newcommand{\EtwoFwNoneChange}{25}
\newcommand{\EtwoFwNoneRate}{3.6\%}

\newcommand{\EtwoFwLaravelShare}{24\%}
\newcommand{\ClusterOneLo}{5.5}
\newcommand{\ClusterOneHi}{8.9}
\newcommand{\ClusterOneN}{2,182}
\newcommand{\ClusterTwoLo}{63.1}
\newcommand{\ClusterTwoHi}{70.7}

\newcommand{\ClusterThreeLo}{12.3\%}
\newcommand{\ClusterThreeHi}{17.9\%}

\newcommand{\EtwoIneligibleLaravelShare}{7.6\%}

\newcommand{\IvRtPatched}{129}

\newcommand{\IvHdSupported}{129}
\newcommand{\IvHdPass}{44}

\newcommand{\IvHdOther}{85}
\newcommand{\IvCarbon}{82}

\newcommand{\IvDoctrine}{3}
\newcommand{\IvTrPairs}{109}
\newcommand{\IvTrBoth}{47}
\newcommand{\IvTrReleaseOnly}{55}

\newcommand{\IvTrNeither}{7}
\newcommand{\EtwoFeasibility}{24}
\newcommand{\EtwoPHPUnitOldest}{4.0}
\newcommand{\EtwoPHPUnitNewest}{9.6}

\newcommand{\EtwoACommitDexist}{14.7\%}

\begin{document}

\title{Yesterday's Code, Today's Dependencies:\\
How Dependency Selection Shapes\\
PHP Migration Benchmarks}


\author{
\IEEEauthorblockN{Afridi Mahmud}
\IEEEauthorblockA{
Bangladesh University of\\
Engineering and Technology\\
Dhaka, Bangladesh\\
\texttt{afridimahmud99@gmail.com}
}
\and
\IEEEauthorblockN{Md. As-Aid Rahman}
\IEEEauthorblockA{
Bangladesh University of\\
Engineering and Technology\\
Dhaka, Bangladesh\\
\texttt{mdasaidrahman@gmail.com}
}
\and
\IEEEauthorblockN{Rifat Shahriyar}
\IEEEauthorblockA{
Bangladesh University of\\
Engineering and Technology\\
Dhaka, Bangladesh\\
\texttt{rifat@cse.buet.ac.bd}
}
}

\maketitle

\begin{abstract}
Migration benchmarks need projects whose tests pass before
migration begins. Yet an old project can behave differently
without any change to its code. Reinstalling it today may
introduce newer libraries, changing both its initial test
results and the failures observed during migration.

We investigate this effect in PHP~7.4-to-8.3 migration.
Using unchanged project source and saved package records,
we compare two dependency policies. Modern selection
allows all versions in the snapshot, while release-date
selection allows only versions published by the project's
release date. With Composer~2 fixed, release-date selection
improves the PHP~7.4 pass rate by \PrimDiff{} across
\PrimN{} releases. A separate comparison against committed
locks shows that it also recovers recorded dependency
versions more closely than modern selection.

The choice matters after migration as well. Among projects
that pass on PHP~7.4 under both policies, PHP~8.3 test
outcomes differ in \EtwoADexist{} of eligible pairs.
We use a shared test runner and bypass Composer's platform
check to isolate test behavior. The resulting candidate
migration task sets overlap by only 58.5\%.
Most differences arise in Laravel projects, where older
dependencies trigger deprecations that newer versions avoid.
Targeted interventions show that removing these initial
failures can restore passing tests or expose further
incompatibilities.

Thus, dependency selection changes not just whether an
old project passes, but also which migration failures a
benchmark can observe. Preserved locks provide the
strongest available version reference; without them,
release-date selection offers a historically grounded
alternative. Migration benchmarks should make this choice
explicit and report its effect on their results.

\end{abstract}

\section{Introduction}\label{sec:intro}

Consider a PHP project that allows any 1.x version of a
library. When released, it installs version~1.2. Years later,
the same requirement may select version~1.9. Both versions
are allowed. Yet the project's tests may produce different
results.

This matters for migration benchmarks. These benchmarks
evaluate tools that move software to newer language versions.
They often select projects whose tests pass on the old version
but fail on the new one~\cite{fujii2026timemachine,liu2026migrationbench,may2026freshbrew}.
A different library version can change whether a project
qualifies for evaluation at all. A lock file preserves the
selected versions~\cite{composer_commands}, but 82.9\% of
releases with a \texttt{composer.json} in our corpus lack one.
For these, a benchmark must choose which dependency versions to install.

Yet the effect of this choice on baseline eligibility and observed
failures remains poorly understood. Reconstruction studies
rebuild old software and treat a successful build or test run
as the goal~\cite{tufano2017compile,maesbermejo2024testing,mukherjee2021pydfix,chan2023rang,malka2024nix}.
Benchmark pipelines filter packages by date or install the
newest versions as a setup decision, but do not report how
this choice changes their task
sets~\cite{zhang2025swebenchlive,vergopoulos2025setupagent,swebench_php_specs}.
TimeMachine-bench selects dependencies by release
date~\cite{fujii2026timemachine}, but its evaluation also
changes the Python version and does not isolate the effect
of dependency selection.

To our knowledge, we are the first to examine this effect in PHP~7.4-to-8.3 migration, keeping
project source unchanged. We compare two policies using
Composer, PHP's package manager, and the same saved
Packagist records. \emph{Modern selection} considers all
versions in the snapshot, while \emph{release-date selection}
considers only versions published by the project's release
date. We hold Composer~2 fixed for the main comparison and
examine Composer version effects separately.

We ask three questions:
\begin{itemize}
    \item \textbf{RQ1: Which projects pass before migration?}
    How do dependency selection and Composer version affect
    test results on PHP~7.4?

    \item \textbf{RQ2: How faithful is the reconstruction?}
    When a preserved lock file exists, how closely do the
    selected versions match it?

    \item \textbf{RQ3: Which failures become visible?}
    For projects passing the PHP~7.4 controls under both
    policies with the same runner, do their PHP~8.3
    outcomes differ?
\end{itemize}

Dependency selection affects the benchmark before migration
even begins. With Composer~2 fixed, release-date selection
raises the PHP~7.4 pass rate by \PrimDiff{} across \PrimN{}
releases (RQ1). In a separate set of releases with committed
locks, it also improves the average share of matching
dependency versions by \EoneDelta{} among releases both
policies resolve (RQ2).

Its influence continues after the PHP upgrade. With
Composer's platform check bypassed, PHP~8.3 pass/fail
outcomes differ in \EtwoADexist{} of eligible pairs, and the
two policies' candidate migration tasks overlap by
58.5\% (RQ3). Nearly all differences favor modern selection,
and most occur in Laravel projects, where an early-deprecation
mechanism dominates. Excluding pairs affected by this
mechanism reduces the observed disagreement rate to
\EtwoMechRestDexist{}.
Preventing the deprecation from stopping tests lets some
projects pass but reveals further failures in others: the
first visible failure can hide another.

Our contribution is a controlled empirical study of how
dependency selection affects three aspects of migration
benchmarks: passing baselines, recovery of recorded versions,
and visible migration failures. We compare candidate task
sets, not tool performance on them. To support reproduction,
we provide a source-archive index, a date-filtered Packagist
mirror, execution procedures, and per-project results.
\section{Background}\label{sec:background}

\subsection{Package selection and lock files}

Composer reads a project's \texttt{composer.json} and
selects package versions that satisfy all requirements,
including those of other packages. These selections and
their relationships form a \emph{dependency graph}.

The command \texttt{composer update} selects package
versions and saves them in \texttt{composer.lock}.
The command \texttt{composer install} follows this lock
when available. Otherwise, it resolves dependencies
again~\cite{composer_commands}. A saved lock tells us
which versions were selected, but not whether the
project's tests passed with them.

\subsection{Composer version compatibility}

Some packages use Composer plugins, which extend the
package manager. Each plugin declares which Composer
interface it supports through
\texttt{composer-plugin-api}~\cite{composer_plugins,composer_platform}.
Composer~1 provides interface version~1.x, while
Composer~2 provides version~2.x~\cite{composer_upgrade2}.

A plugin requiring \texttt{composer-plugin-api \^{}1.0}
can therefore block installation with Composer~2.
Disabling plugins does not remove this requirement.
An installation may fail because of the Composer version,
even when the required library versions are available.

\subsection{Changes from PHP 7.4 to PHP 8.3}

PHP~8 introduced changes that can break older code.
For example, PHP~8.0 removed functions such as
\texttt{each()} and
\texttt{create\_function()}~\cite{php80_incompatible}.

PHP~8.1 also introduced deprecation notices for methods
that override certain built-in methods without compatible
return types~\cite{php81_incompatible}.
These notices warn about future incompatibility.
They do not stop execution by themselves. However, a
project's error handler may turn them into exceptions
that stop its tests. Thus, a PHP~8.3 failure can depend
on both the code and how the project handles warnings.

\section{Study Design}\label{sec:design}

\begin{figure*}[t]
\centering
\includegraphics[width=\textwidth]{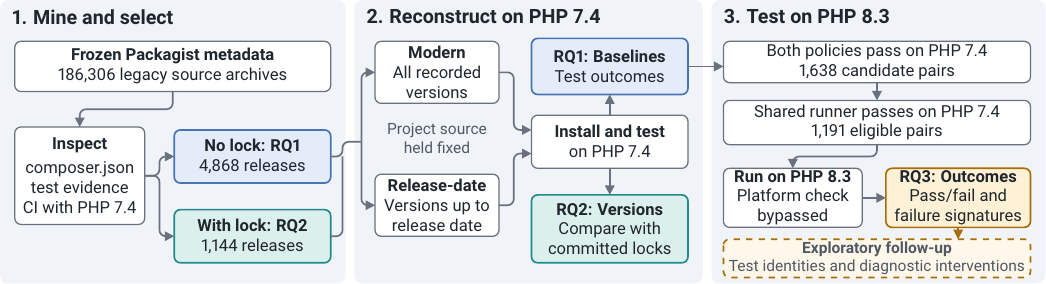}
\caption{Study overview. RQ1 measures passing baselines
using releases without locks. RQ2 checks version recovery
in a separate lock-based set. RQ3 follows RQ1 releases
that pass under both policies and uses a shared test runner
before comparing PHP~8.3 outcomes. The dashed box marks
exploratory checks.}
\label{fig:overview}
\end{figure*}

We compare installations of the same project release under two
dependency-selection policies. Within each primary comparison,
we fix the source archive, dependency requirements, registry
snapshot, PHP version, and Composer version; only the dependency
versions selected by Composer may differ. Development dependencies,
including PHPUnit, can also change, so RQ3 uses a shared
external test runner. We examine three consequences
(Figure~\ref{fig:overview}):

\begin{itemize}
\item \textbf{RQ1, baselines:} Which releases pass on PHP~7.4
under each policy? Main measure: the paired difference
in pass rates.
\item \textbf{RQ2, fidelity:} How closely does each policy
recover versions recorded in committed locks? Main measure:
the share of locked versions recovered.
\item \textbf{RQ3, failures:} Do the two installations behave
differently on PHP~8.3? Main measure: the share of pairs
with different pass/fail outcomes.
\end{itemize}

Signed differences are release-date minus modern selection,
so positive values favor release-date selection.

\providecommand{\rqanswer}[2]{\par\medskip\noindent\fbox{\parbox{\dimexpr\columnwidth-2\fboxsep-2\fboxrule\relax}{%
\textbf{Answer to #1:} #2}}\par\medskip}

\subsection{Dependency policies}
\label{sec:design-pipeline}\label{sec:design-arms}

Both policies use the same release archive, dependency
requirements, and saved Packagist records.
\emph{Modern selection} permits every version in the
snapshot, as a fresh installation would.
\emph{Release-date selection} permits only versions
published by the project's release date. Both preserve
declared PHP and extension constraints.

The primary comparison uses Composer~2 (\ComposerTwo{})
under both policies: A2 for modern selection and B2 for
release-date selection. Since Composer~2 appeared in
October 2020, after most study releases, we repeat the
comparison with Composer~1 (\ComposerOne{}), which some
older plugins require. A \emph{historical recovery}
setting tries release-date selection with Composer~2,
then falls back to Composer~1 only if installation fails.
This rule was fixed before execution and does not depend
on test results. Table~\ref{tab:arms} lists all settings.

\begin{table}[t]
\centering
\caption{Dependency settings and the codes used in figures and replication records.}
\label{tab:arms}
\small
\begin{tabular}{@{}llll@{}}
\toprule
Code & Selection & Composer & Used in \\
\midrule
A2 & Modern & 2 & All RQs \\
B2 & Release date & 2 & All RQs \\
A1 & Modern & 1 & RQ1 \\
B1 & Release date & 1 & RQ1 \\
HR & Release date & 2, else 1 & RQ1 \\
L & Committed lock & 2 & RQ2 \\
P & Lowest allowed & 2 & RQ2 \\
B2L & Lock-informed date & 2 & RQ2 \\
\bottomrule
\end{tabular}
\end{table}

\subsection{Study sets}\label{sec:design-frame}

For each package, we select its latest stable release
available by 25 November 2020, the day before PHP~8.0
was released. We treat each selected release as one
observation (Section~\ref{sec:corpus}).
Table~\ref{tab:studysets} shows the selection rules
and main-analysis subsets.

\begin{table*}[t]
\centering
\caption{Study sets and the subsets used in each main analysis.}
\label{tab:studysets}
\small
\begin{tabular}{@{}llrlr@{}}
\toprule
RQ & Selection rule & Selected & Main-analysis criterion & Main set \\
\midrule
RQ1 & No lock; tests; PHP~7.4 CI job & \NFrame{}
  & No infrastructure failure; runnable under $\geq$1 policy & \PrimN{} \\
RQ2 & Lock; otherwise as RQ1 & \EoneFrame{}
  & High-confidence lock; both policies resolve & \EoneToneN{} \\
RQ3 & RQ1 passes under both policies & \EtwoPairs{}
  & Shared runner passes on PHP~7.4, starts on 8.3 & \EtwoAEligible{} \\
\bottomrule
\end{tabular}
\end{table*}

\textbf{RQ1} requires \texttt{composer.json}, test
evidence, no committed lock, and a Travis~CI or GitHub
Actions configuration with a PHP~7.4 job. CI indicates
intended support, not historical success, so we execute
the tests ourselves. Releases are processed in random
order (seed \FrameSeed{}).

\textbf{RQ2} uses separate releases meeting the same
criteria but with committed locks recording installed
versions. Of \EoneCandidates{}, we exclude
\EoneExcludedRepos{} with unsupported repositories and
\EoneUnparseable{} with unreadable locks. We form three
nested groups to assess lock quality:
\emph{high-confidence} locks (\EoneTone{}, primary)
agree with \texttt{composer.json}, pass Composer
validation, and appear to be installed as-is in CI,
based on configuration patterns; \emph{current} locks
(\EoneTtwo{}) need only agree with \texttt{composer.json};
and \emph{all} locks (\EoneTthree{}) include every release.

\textbf{RQ3} uses RQ1 releases that pass on PHP~7.4 under
both Composer~2 policies. The two installations of each
release form a \emph{pair}. We set aside
\EtwoFeasibility{} pairs for a feasibility check.

\subsection{Execution and outcomes}\label{sec:design-exec}

For baseline evaluation, we run each archive in a
PHP~\PHPSeven{} container with Composer scripts and
plugins disabled. Metadata comes only from our mirror;
containers cannot reach Packagist but download archives
from hosting sites. We use \texttt{composer update} to
reconstruct dependencies under each policy and
\texttt{composer install} for committed-lock installations.
We then run tests with a 900-second limit.
Infrastructure failures, such as container or download
errors, are recorded separately and retried twice.
Inconsistent outcomes trigger two further runs;
we take the majority of three.

An installation is \emph{runnable} if dependencies
install and the test command executes, regardless of
the outcome. It \emph{passes} if the test command exits
with status zero and reports no failures or errors.
When several problems occur, we record the first in
this order: platform rejection, timeout, PHP fatal
error, no JUnit report, no tests in the report, pass,
failure. PHPUnit may report risky tests as errors
despite exit status zero. All-skipped suites count
as passing; Table~\ref{tab:sensitivity} checks this choice.

\subsection{RQ1: baselines on PHP~7.4}
\label{sec:design-analysis}

RQ1 asks whether dependency selection changes which
releases pass on PHP~7.4.

\textbf{Main measure.} We compare the same releases
under both policies. A \emph{gain} is a release that
passes only under release-date selection; a \emph{loss}
passes only under modern selection. The paired pass-rate
difference equals gains minus losses, divided by the
number of releases. A release enters a comparison only
if neither setting has an infrastructure failure and
at least one installation is runnable. Thus, denominators
may differ across comparisons.

\textbf{Statistics.} An exact McNemar test compares gains
and losses; a Newcombe interval~\cite{newcombe1998paired}
bounds the paired difference. We also report the
\emph{recoverable-rejection rate}: among releases with
runnable modern tests, the share that fail under modern
selection but pass under historical recovery (Wilson
interval). A rescued release does not establish that
the modern failure was spurious.

\textbf{Tracing failures.}\label{sec:design-attribution}
To investigate why gained releases fail under modern
selection, we start from their release-date installations
and ask Composer (\texttt{update -W --minimal-changes})
to upgrade named packages under modern selection.
Delta debugging~\cite{zeller2002ddmin} seeks a smaller
request that still fails; uninstallable requests do
not reproduce the failure. A request is \emph{1-minimal}
if removing any package restores passing tests or
prevents installation. Searches stop after 40
evaluations (checked between iterations), and fresh
runs confirm each failure. Composer may change
unrequested packages, so a request identifies a
failure trigger, not a defective library.

\subsection{RQ2: fidelity to committed locks}
\label{sec:design-fidelity}

RQ2 asks how closely the policies reproduce recorded
versions, regardless of test outcomes.

\textbf{Main measure.} The \emph{recovery share} is the
fraction of a committed lock's packages, including
development dependencies, reproduced at the same
version. We ignore a leading \texttt{v}, require
development versions to match the source commit,
and count missing packages as mismatches. For each
release, the paired difference is the release-date
share minus the modern share.

\textbf{Statistics.} The primary analysis uses
high-confidence locks resolved by both policies,
with an exact sign test and a bootstrap interval
from 10,000 release-level resamples. Secondary
analyses use broader lock groups, count unresolved
releases as zero recovery, and compare tests against
committed-lock installations, with Holm adjustment.

\textbf{Alternative settings.} To check whether
simply choosing older versions explains the result,
\emph{lowest-allowed} selection picks the oldest
permitted versions. \emph{Lock-informed} selection
sets the cutoff to the newest locked package's date.
Because it requires the lock, this is a diagnostic
only and runs no tests.

\subsection{RQ3: failures on PHP~8.3}
\label{sec:design-distortion}

RQ3 asks whether the two installations behave
differently after the PHP upgrade.

\textbf{Fixed runner.} For each pair, we preserve both
dependency graphs and change only PHP~7.4 to PHP~8.3.
The installations use PHPUnit versions ranging from
\EtwoPHPUnitOldest{} to \EtwoPHPUnitNewest{}, which
could confound library effects with runner effects.
We therefore use one external runner for both: a
self-contained PHPUnit~8.5.55 PHAR, with 9.6.37
as fallback. A pair is \emph{eligible}
if this runner passes both installations on PHP~7.4
and starts on PHP~8.3.

\textbf{Platform check.} Composer may reject PHP~8.3
because dependencies do not declare support. We
bypass this check and record its verdict separately;
passing tests therefore show behavior under PHP~8.3,
not migration readiness.

\textbf{Outcome categories.} We classify each eligible
pair by its PHP~8.3 outcomes: \emph{both pass} if both
installations pass, \emph{masked} if only modern selection
passes, or \emph{manufactured} if only release-date
selection passes. If both fail, the pair is \emph{stable}
when signatures match or \emph{transformed} when they
differ. The names describe modern selection: it masks
a failure visible under release-date selection or
manufactures one absent there. A \emph{signature} contains
normalized failing-test names and count, or a fatal
error's message and code location. For example,
\texttt{ambersive/}\allowbreak\texttt{documentviewer}
passes on PHP~7.4 under both policies. On PHP~8.3, its
modern installation passes, while its release-date
installation stops on a Symfony deprecation, making
it a \emph{masked} pair.

\textbf{Main measure.} The \emph{outcome-change rate}
is the share of eligible pairs that are masked or
manufactured. An exact McNemar test compares their counts.
A broader rate includes transformed pairs, although
different signatures may not require different repairs.
Since all eligible pairs pass on PHP~7.4, a pair becomes
a \emph{candidate migration task} under a policy when
its PHP~8.3 tests fail. These are test-defined candidates,
not necessarily platform-compatible installations.
Inference treats releases as independent
(Section~\ref{sec:threats}).

\textbf{Further checks.} Robustness checks use PHPUnit
installed through Composer, which may change other
dependencies, and each project's own runner. Exploratory
interventions target PHP~8.1 return-type deprecations:
we add \texttt{ReturnTypeWillChange} attributes to
affected dependency methods or change Laravel's error
handler so deprecations no longer stop tests, then
check whether tests pass or another failure appears.

\subsection{Label checks}\label{sec:design-manual}

Our main results use tests and locks, not model judgments.
To check interpretations, three agents (GPT-6 Astra,
Claude Opus~5.5, and Gemini~3.1 Pro) inspect \ManualN{}
randomly drawn RQ1 gains and \EtwoManualN{} RQ3 pairs
with changed outcomes or signatures, stratified by
category. They see only relevant evidence, with no web
access or follow-up prompts; RQ3 identities and
categories are hidden.

After refining the coding guide, each agent codes
fresh samples (\ValRqOneN{} RQ1 and
\ValRqThreeN{} RQ3) twice, with \ProbeN{} planted
errors included. We report majority support and Fleiss'
$\kappa$ with bootstrap intervals~\cite{ahmed2025llm};
labels without majority support remain unvalidated.
Agent agreement is not a human reference standard.

\subsection{Decision timing}\label{sec:design-prereg}

RQ1's study set and comparisons were fixed before
execution. RQ2 and RQ3 were designed after RQ1
results but before their own runs. After inspecting
RQ3 outcomes, we made pass/fail changes the primary
measure instead of all signature changes; both are
reported. We also added framework analyses,
comparability checks, and interventions as exploratory
analyses. The replication package records these
decisions and execution details.

\section{Corpus and Mirror}\label{sec:corpus}

\subsection{Release collection}\label{sec:corpus-funnel}

We collected the Packagist package listing on \ListedOn{}
and retrieved available version records, including release
dates. Records for \NGone{} listed packages were no longer
available. We then downloaded and indexed the qualifying
legacy release archives (Section~\ref{sec:design-frame}).

Table~\ref{tab:funnel} shows how the collection narrows
to the RQ1 and RQ2 study sets. We preserve the source
archives so later repository changes cannot affect
the releases used in our experiments.

\begin{table}[t]
\centering
\caption{Corpus collection and study-set selection.
The first two rows count packages; the rest count releases.}
\label{tab:funnel}
\begin{tabular}{@{}lrr@{}}
\toprule
Collection step & \multicolumn{2}{r@{}}{Count} \\
\midrule
Packages listed & \multicolumn{2}{r@{}}{\NPackagist} \\
Metadata retrieved & \multicolumn{2}{r@{}}{\NRetrieved} \\
Legacy releases & \multicolumn{2}{r@{}}{\NLegacy} \\
Archives indexed & \multicolumn{2}{r@{}}{\NIndexed} \\
With \texttt{composer.json} &
    \multicolumn{2}{r@{}}{\NComposerJson} \\
\midrule
Selection step & No lock & With lock \\
\midrule
Lock-file status & \NLockless & \EoneWithLock \\
With test evidence & \NTests & \EoneLockTests \\
CI includes PHP~7.4 & \textbf{\NFrame} & \EoneCandidates \\
After RQ2 exclusions & & \textbf{\EoneFrame} \\
\bottomrule
\end{tabular}
\end{table}

\subsection{Date-filtered package mirror}\label{sec:corpus-mirror}

We built a local mirror to supply both policies with the
same saved package records. It filters versions by their
recorded release dates without contacting live Packagist.

Two implementation details are important.
First, Packagist compresses some Composer~2 metadata by
storing differences between version records
~\cite{composer_repositories,composer_minifier}.
An older version may depend on fields stored with a newer
one. We expand these records before applying the date
cutoff. This preserves dependency requirements even when
newer versions are excluded.

Second, the mirror supports both Composer~1 and
Composer~2 repository formats~\cite{composer_repositories}.
For Composer~1, we generate version identifiers and index
packages that provide or replace other package names.
This allows Composer to resolve dependencies through
those alternative names.

The mirror preserves the records available at collection
time. It cannot recover missing versions or development
branches without release dates. Projects requiring these
records may therefore remain unresolved under either
policy.

\section{RQ1: Which Projects Pass Before Migration?}
\label{sec:results}

A project's dependency versions can determine whether it qualifies as a passing baseline. On PHP~7.4, release-date
selection raises the paired pass rate by \PrimDiff{} with Composer~2 fixed. Allowing a Composer~1 fallback increases
the difference to \PracDiff{}. The releases gained by release-date selection also install under modern selection;
what changes is whether their tests pass with the selected libraries.

\subsection{Release-date selection preserves more passing baselines}
\label{sec:results-validity}

Of \NProcessed{} selected releases, \NDeclaresRepos{} use repositories our mirror cannot serve, leaving
\NWithOutcomes{} with recorded outcomes. We exclude another 1,038 because at least one installation has an
infrastructure failure or neither policy produces a runnable test. The primary paired comparison therefore contains
\PrimN{} releases, with PHP~7.4 and Composer~2 held fixed.

With modern selection, 47.8\% of these releases pass, compared with 54.9\% under release-date selection. The paired
gain is \PrimDiff{} (95\% CI \PrimDiffLo{}--\PrimDiffHi{}\,pp; exact McNemar \PrimP{}). This reflects
\PrimRecovered{} releases that pass only under release-date selection and \PrimRegressed{} that pass only under
modern selection (Figure~\ref{fig:forest}, top row).

These gains are not simply successful installations. All \Recoveries{} gained releases also install under modern
selection, but their tests do not pass: \RecovFail{} report test failures, \RecovFatal{} encounter fatal errors,
and \RecovNoReport{} produce no usable report. The same release can therefore enter a migration benchmark as a
passing baseline under one policy but not the other.

\subsection{The Composer version matters too}
\label{sec:results-decomposition}

\begin{figure}[t]
\centering
\includegraphics[width=\columnwidth]{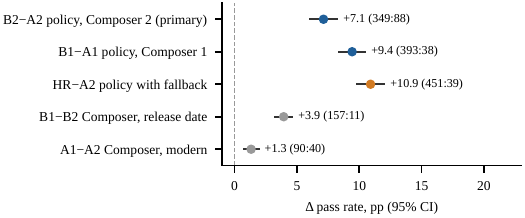}
\caption{Paired pass-rate differences with 95\% confidence intervals. The top row is the primary comparison; the
others separate dependency selection, Composer version, and the pre-set fallback. Parentheses show passing baselines
gained and lost by the first setting.}
\label{fig:forest}
\end{figure}

Dependency selection is not the only choice that affects these baselines. Composer's version also matters,
particularly for older plugins. Figure~\ref{fig:forest} separates the two effects. Each paired comparison excludes
infrastructure failures under either setting, so its population ranges from \PrimN{} to \PracN{} releases.

\emph{Selection policy.} With Composer~1 fixed, release-date selection still increases the paired pass rate by
\MetaOneDiff{} across \MetaOneN{} releases.

\emph{Composer version.} Switching from Composer~2 to Composer~1 adds \ToolDiff{} under release-date selection,
compared with \ToolModernDiff{} under modern selection. The difference largely involves plugins: among the
\ToolRecovered{} release-date installations that pass only with Composer~1, \DescBoneGainsPlugin{} use plugins
requiring the older interface. Plugin requirements also explain many primary-comparison losses. Of the
\PrimRegressed{} releases passing only under modern selection, \DescBtwoLossesPlugin{} involve older plugin
requirements; a Composer~1 fallback recovers \DescBtwoLossesRescued{} of them.

\emph{Historical recovery.} We then combine release-date selection with a Composer~1 fallback, using the rule fixed
before execution and independent of test outcomes. Relative to modern selection with Composer~2, this increases the
paired pass rate by \PracDiff{}. Among releases with runnable modern tests, \FalseBaseNum{} of \FalseBaseDen{}
(\FalseBaseShare{}; 95\% Wilson CI \FalseBaseLo{}--\FalseBaseHi{}) fail under modern selection but pass under
historical recovery. Unlike the fixed-Composer comparison, this result combines dependency selection with
package-manager compatibility. Section~\ref{sec:threats} describes a correction to the Composer~1 results.

\subsection{How dependency upgrades change test results}
\label{sec:results-drift}\label{sec:mechanisms}

The two policies usually select different dependencies. Among \DriftCompared{} releases with locks from both
Composer~2 installations, \DriftDifferShare{} have different dependency graphs. The median release differs in
\DriftChangedMedian{} packages; in \DriftMajorShare{}, at least one shared package changes its major version, or
its minor version below~1.0.

To examine whether these upgrades trigger failures, we start from passing release-date installations, request
modern versions of selected packages, and rerun the tests (Section~\ref{sec:design-attribution}). Delta debugging
finds a 1-minimal failure-inducing request for \AttrAttributed{} of the \Recoveries{} gained releases. In
\AttrCoMoved{} cases, Composer also changes unrequested packages, so a request identifies a trigger rather than a
single defective library.

We then check whether the failures reproduce. Fresh runs reapply \ConfCases{} requests: \AttrAttributed{} from
completed searches and \ConfPartial{} from searches that reached the stopping threshold. In \ConfSignature{} cases,
the reproduced failure shares a failing test or fatal message with the modern installation. Thus, permitted upgrades
reproduce the observed failures in most cases, although matching errors do not prove a single cause.

Agent judgments provide a separate, more limited check. LLM agents implicate the requested upgrades in
\ManualImplicated{} of \ManualN{} original cases and \ValCulpritYes{} of \ValRqOneN{} fresh cases. Agreement is high
in the original sample ($\kappa=\ManualKappaCulprit{}$) and lower in the fresh sample
($\kappa=\ValKappaCulprit{}$; 95\% CI \ValKappaCulpritLo{}--\ValKappaCulpritHi{}), where nearly all judgments are
positive, which depresses $\kappa$. These judgments support the interpretation but do not provide human-verified
causal labels.

\subsection{How reliable is the improvement?}
\label{sec:results-robust}

The gain remains positive across the checks in Table~\ref{tab:sensitivity}, from \SensCommitDiff{} when each source
commit counts once to \SensExecDiff{} when a usable test report is required. Exploratory resampling of
\ClusterOneN{} vendor groups also gives a positive interval (\ClusterOneLo{}--\ClusterOneHi{}\,pp). Among
\FlakyRerun{} targeted repeated executions, \FlakyN{} change outcome; these runs do not estimate flakiness across
the whole corpus.

\begin{table}[t]
\centering
\caption{Robustness of the paired Composer~2 comparison.
Gains pass only with release-date selection; losses
only with modern selection.}
\label{tab:sensitivity}
\setlength{\tabcolsep}{4.5pt}
\begin{tabular}{@{}lrrrl@{}}
\toprule
Analysis set & $n$ & Gained & Lost & Difference [95\% CI] \\
\midrule
Primary &
\PrimN & \PrimRecovered & \PrimRegressed &
\PrimDiff{} [\PrimDiffLo, \PrimDiffHi] \\
Consistent reruns &
\SensStableN & \SensStableRecovered & \SensStableRegressed &
\SensStableDiff{} [\SensStableDiffLo, \SensStableDiffHi] \\
At least one report &
\SensExecN & \SensExecRecovered & \SensExecRegressed &
\SensExecDiff{} [\SensExecDiffLo, \SensExecDiffHi] \\
No all-skipped passes &
\SensSkipN & \SensSkipRecovered & \SensSkipRegressed &
\SensSkipDiff{} [\SensSkipDiffLo, \SensSkipDiffHi] \\
One per source commit &
\SensCommitN & \SensCommitRecovered & \SensCommitRegressed &
\SensCommitDiff{} [\SensCommitDiffLo, \SensCommitDiffHi] \\
\bottomrule
\end{tabular}
\end{table}

\rqanswer{RQ1}{Release-date selection raises the PHP~7.4 pass rate by \PrimDiff{} with Composer~2 fixed and by
\PracDiff{} with a Composer~1 fallback. Upgrades permitted by the projects' own constraints trigger most lost
baselines, so dependency selection decides which projects can enter a migration benchmark.}

\section{RQ2: How Closely Are Recorded Versions Recovered?}
\label{sec:fidelity}

Passing tests do not establish that a reconstruction matches the dependency versions developers recorded. We therefore
compare reconstructed versions with committed locks from a separate set of releases. Among the releases each policy
resolves, release-date selection recovers \EoneMBtwo{} of locked versions on average, compared with \EoneMAtwo{} for
modern selection.

\subsection{Release dates recover more recorded versions}

The primary group contains \EoneTone{} releases whose current locks are installed by CI. One lock is empty,
leaving \EoneGraphTone{}. Table~\ref{tab:recovery} summarizes recovery. Release-date and modern selection also
have \EoneInfraBtwo{} and \EoneInfraAtwo{} infrastructure failures, respectively.

\begin{table}[t]
\centering
\caption{Version recovery on the \EoneGraphTone{} high-confidence locks. Mean: average recovery share over the releases each setting resolves; Complete: locks reproduced exactly. Unresolved counts exclude infrastructure failures.
Lock-informed date selection is exploratory and runs no tests.}
\label{tab:recovery}
\footnotesize
\setlength{\tabcolsep}{4pt}
\begin{tabular}{@{}lrrrr@{}}
\toprule
Setting & Resolved & Unresolved & Mean & Complete \\
\midrule
Release date & \EoneLockBtwo{} & \EoneUnresBtwo{} & \EoneMBtwo{} & \EoneExactBtwo{} \\
Modern & \EoneLockAtwo{} & \EoneUnresAtwo{} & \EoneMAtwo{} & \EoneExactAtwo{} \\
Lowest allowed & \EoneLockP{} & \EoneUnresP{} & \EoneMP{} & \EoneExactP{} \\
Lock-informed date & \EoneLockBL{} & \EoneUnresBL{} & \EoneMBL{} & \EoneExactBL{} \\
\bottomrule
\end{tabular}
\end{table}

The means in Table~\ref{tab:recovery} cover different sets of resolved releases. For a direct comparison, we use the
\EoneHN{} releases that both policies resolve. Release-date selection recovers more locked versions in
\EoneHCloser{} of these, fewer in \EoneHFurther{}, and the same share in \EoneHTies{}. Its mean paired advantage
is \EoneDelta{} (95\% bootstrap interval \EoneDeltaLo{}--\EoneDeltaHi{}\,pp; exact sign test \EoneHP{}). Across their
respective resolved releases, release-date selection reproduces \EoneExactBtwo{} locks completely, compared with
\EoneExactAtwo{} for modern selection.

\begin{figure}[!b]
\centering
\includegraphics[width=\columnwidth]{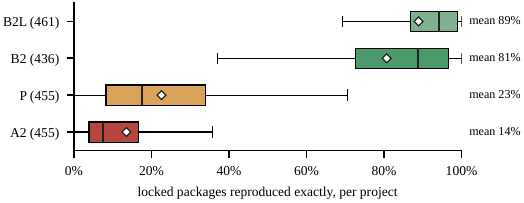}
\caption{Share of locked versions recovered per release. Boxes show quartiles; diamonds show means. Parentheses give
the number of releases each setting resolves, so the boxes cover different releases. A2: modern; B2: release date;
P: lowest allowed versions; B2L: cutoff inferred from the lock.}
\label{fig:e1}
\end{figure}

We check whether choosing older versions or omitting unresolved installations explains the advantage.
Lowest-allowed selection recovers only \EoneMP{} on average (Figure~\ref{fig:e1}), well below release-date
selection. When unresolved reconstructions count as recovering nothing, release-date selection still does better in
\EoneSensCloser{} of \EoneSensN{} releases without infrastructure failures under either policy and worse in
\EoneSensFurther{}, with a paired mean advantage of \EoneSensDelta{}.

\subsection{Why recovery remains imperfect}

Most remaining mismatches involve versions that are too new. A lock can predate the project's release, so a cutoff at
the release date may admit versions published after the lock was written. Among mismatches with comparable release
times, \EoneNewerBtwo{} selected versions are newer than the locked ones, while only \EoneOlderBtwo{} are older. Some
versions are also unavailable: \EoneAbsentProjects{} releases lock a package or version missing from the mirror.

Using the newest locked package's date as the cutoff improves recovery in \EoneBLCloser{} releases and reduces it
in \EoneBLFurther{}, with a mean recovery of \EoneMBL{}. This shows that cutoff timing matters, but the method
needs the lock and cannot help when none exists.

\subsection{Recovered versions must also run}

Recovering versions does not guarantee passing tests. We therefore compare reconstructions with
committed-lock installations. Among the \EoneFBBtwoDen{} releases whose committed-lock
installation passes, reconstruction fails to produce a pass in \EoneFBBtwoShare{} of cases under release-date
selection, \EoneFBAtwoShare{} under modern selection, and \EoneFBPShare{} under lowest-allowed selection. Most
release-date failures occur before tests begin: \EoneFBBtwoUnresolvable{} of \EoneFBBtwo{} cannot resolve dependencies.
In \EoneSoneN{} paired comparisons, \EoneSoneBtwo{} releases pass only with release-date selection and none only with
modern selection (Holm-adjusted \EoneSoneP{}).

As a measurement check, the lock saved after each committed-lock installation matches the committed lock in all
\EoneLLocks{} checked cases. This validates version comparison, not whether the tests originally passed.

\subsection{Robustness checks}

The advantage persists in the broader lock groups: \EoneTtwoDelta{} for current locks (\EoneTtwoN{} paired
releases) and \EoneTthreeDelta{} for all locks (\EoneTthreeN{}). It is \EoneCommitDelta{} when each source commit
counts once, and exploratory vendor-level resampling gives a positive interval (\ClusterTwoLo{}--\ClusterTwoHi{}\,pp).

These checks support the result, but not its direct generalization to RQ1. The lock-based sample resembles RQ1
in release year (\EoneYearShare{} versus \MainYearShare{} from 2020) but contains more libraries
(\EoneLibShare{} versus \MainLibShare{}). Together, RQ1 and RQ2 show that release-date selection preserves more
passing baselines and comes closer to the versions developers recorded.

\rqanswer{RQ2}{Release-date selection reproduces \EoneMBtwo{} of locked versions on average, against \EoneMAtwo{} for
modern selection, and is closer in \EoneHCloser{} of \EoneHN{} paired releases. Most remaining mismatches select
versions newer than the lock, because locks can predate the release.}

\section{RQ3: Which Failures Appear After Migration?}
\label{sec:distortion}

On PHP~8.3, the two policies disagree on whether tests pass in \EtwoADexist{} of eligible pairs, almost always because
modern libraries pass where release-date libraries fail. One framework-specific mechanism explains most of this
difference, and removing it often exposes a second failure.

\subsection{The same projects, different outcomes}

RQ3 starts from projects that pass on PHP~7.4 under both policies. We keep each installation's libraries fixed and run
both on PHP~8.3 with the same external test runner (Section~\ref{sec:design-distortion}). Of \EtwoPairs{} pairs,
\EtwoAEligible{} have a runner that works for both installations; most excluded pairs lack one that passes the PHP~7.4
controls. Laravel projects are more common among eligible pairs than excluded pairs (\EtwoFwLaravelShare{} versus
\EtwoIneligibleLaravelShare{}). Our estimates therefore apply to this eligible population.

\begin{figure}[!b]
\centering
\includegraphics[width=\columnwidth]{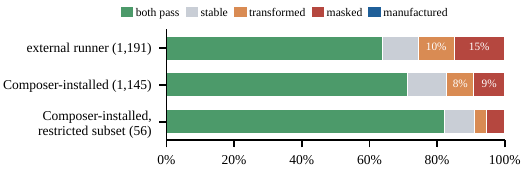}
\caption{PHP~8.3 outcomes under the two dependency policies. The top row uses a fixed external runner. The middle row
uses Composer-installed runners, which may change dependencies. The bottom row keeps only pairs whose updates stay
within runner packages and whose PHPUnit versions match. Counts show eligible pairs. \emph{Both pass}: neither fails;
\emph{stable}: both fail with the same signature; \emph{transformed}: both fail with different signatures;
\emph{masked}: only modern selection passes; \emph{manufactured}: only release-date selection passes.}
\label{fig:e2}
\end{figure}

The installations disagree on whether tests pass in \EtwoADexistN{} pairs, or \EtwoADexist{} (95\% CI
\EtwoADexistLo{}--\EtwoADexistHi{}; Figure~\ref{fig:e2}, top row). In \EtwoAMasked{} of them, modern libraries pass
and release-date libraries fail; only \EtwoAManufactured{} pair shows the reverse (exact McNemar \EtwoAP{}). This
direction is expected, since newer versions are more likely to support PHP~8.3; what matters for benchmarks is the
size of the difference and its cause. In every outcome-changing pair, the PHP~7.4 reports show both installations
running the same test cases and skips.

Composer's PHP~8.3 compatibility check rejects all \EtwoSfiveBtwoAll{} release-date installations, compared with
\EtwoSfiveBoth{} modern installations. We bypass it to compare test behavior under PHP~8.3; with the check enforced,
no release-date installation would reach its tests. Passing under this protocol therefore does not establish migration
readiness.

\subsection{Which releases qualify as candidate tasks?}
\label{sec:rq3-tasksets}

These outcome differences also affect candidate task selection. Migration benchmarks commonly choose projects that
pass on the old PHP version but fail on the new one. All eligible pairs pass on PHP~7.4, so they qualify under a
policy if their PHP~8.3 tests do not pass. Release-date selection yields \EtwoAFailingBtwo{} candidate tasks and
modern selection \EtwoAFailingAtwo{} (Table~\ref{tab:tasksets}).
Only \EtwoABothFail{} are shared, a Jaccard overlap of 58.5\%; the remaining \EtwoANoTask{} pairs pass under both
policies.
These are test-defined candidates with the platform check bypassed, not necessarily installable projects on PHP~8.3.
We do not evaluate their effect on migration-tool scores.

\begin{table}[t]
\centering
\caption{Candidate migration tasks among the \EtwoAEligible{} eligible pairs: pairs that pass on PHP~7.4 but not on
PHP~8.3 under a policy (fixed runner, platform check bypassed).}
\label{tab:tasksets}
\footnotesize
\begin{tabular}{@{}lr@{}}
\toprule
Task-set membership & Pairs \\
\midrule
Release-date selection only (masked) & \EtwoAMasked{} \\
Both policies (stable or transformed) & \EtwoABothFail{} \\
Modern selection only (manufactured) & \EtwoAManufactured{} \\
Neither (both pass on PHP~8.3) & \EtwoANoTask{} \\
\midrule
Total under release-date selection & \EtwoAFailingBtwo{} \\
Total under modern selection & \EtwoAFailingAtwo{} \\
\bottomrule
\end{tabular}
\end{table}

\subsection{Why the differences concentrate in Laravel}

Most masked pairs share one failure. In \EtwoMechMaskedDepr{} of the \EtwoAMasked{}, the release-date installation
stops on a PHP~8.1 return-type deprecation. Deprecations are warnings and do not normally stop execution, but a
framework error handler converts them into exceptions. Laravel provides that handler in \EtwoMechLaravel{} cases, and
the warning comes from \texttt{symfony/http-foundation} in \EtwoMechHttpFoundation{}. The external runner does not
escalate deprecations itself.

As a result, outcome changes cluster in Laravel projects (Table~\ref{tab:rq3-framework}). Laravel-related pairs
account for \EtwoFwLaravelChange{} of the \EtwoADexistN{} changes but only \EtwoFwLaravelShare{} of eligible pairs.
Excluding the \EtwoMechMaskedDepr{} early-deprecation pairs leaves 49 changes among \EtwoMechRestN{} pairs
(\EtwoMechRestDexist{}; 95\% CI \EtwoMechRestDexistLo{}--\EtwoMechRestDexistHi{}).
The overall disagreement rate is therefore strongly shaped by Laravel-related projects and this early-deprecation
mechanism. Both the framework grouping and the exclusion are exploratory.

\begin{table}[!ht]
\centering
\caption{PHP~8.3 outcome changes by framework indicator (exploratory). Groups are disjoint.}
\label{tab:rq3-framework}
\footnotesize
\setlength{\tabcolsep}{3pt}
\begin{tabular}{@{}lrr@{}}
\toprule
Framework indicator & Changed / pairs & Rate \\
\midrule
Laravel-related & \EtwoFwLaravelChange{} / \EtwoFwLaravelN{} & \EtwoFwLaravelRate{} \\
Other identified & 23 / 205 & 11.2\% \\ 
None identified & \EtwoFwNoneChange{} / \EtwoFwNoneN{} & \EtwoFwNoneRate{} \\
\midrule
All eligible & \EtwoADexistN{} / \EtwoAEligible{} & \EtwoADexist{} \\
\bottomrule
\end{tabular}
\end{table}

\subsection{What happens after the first failure?}

The deprecation marks where the tests stopped, not necessarily everything that is wrong. Two exploratory
interventions on fresh installations, with the same runner and platform settings, test what lies behind it
(Table~\ref{tab:interventions}). Adding \texttt{ReturnTypeWillChange} attributes to the affected dependency methods
makes no installation pass, because further deprecations appear. Changing Laravel's error handler so that
deprecations no longer stop tests, while other errors are handled normally, lets \IvHdPass{} of \IvHdSupported{}
release-date installations pass. The other \IvHdOther{} run further and then stop on a Carbon type error
(\IvCarbon{}) or a Doctrine method incompatibility (\IvDoctrine{}).

\begin{table}[!b]
\centering
\caption{Exploratory interventions on PHP~8.3. The runner is fixed and Composer's platform check is bypassed.}
\label{tab:interventions}
\footnotesize
\begin{tabular}{@{}p{.48\columnwidth}p{.43\columnwidth}@{}}
\toprule
Intervention & Observed result \\
\midrule
Add return-type attributes (\IvRtPatched{} installations)
& None pass; further deprecations appear. \\
\addlinespace
Change Laravel's error handler (\IvHdSupported{} installations)
& \IvHdPass{} pass; \IvHdOther{} expose other failures. \\
\addlinespace
Change the handler under both policies (\IvTrPairs{} transformed pairs with the early deprecation;
Section~\ref{sec:results-transformed})
& \IvTrBoth{} pass under both; \IvTrReleaseOnly{} pass only with modern selection; \IvTrNeither{} pass under
neither. \\
\bottomrule
\end{tabular}
\end{table}

For example, \texttt{ambersive/}\allowbreak\texttt{documentviewer} first fails on a Symfony deprecation. With the
handler changed, it runs until a Carbon type error, while its modern installation passes unchanged. Restoring the
original files reproduces each initial failure, and every validated pass runs the same test cases, including skips,
as the modern installation. Removing the first obstacle can thus restore passing tests or reveal another
incompatibility.

\subsection{Different errors can hide the same problem}
\label{sec:results-transformed}

When both installations fail, their errors often differ. In \EtwoATransformed{} pairs, both fail with different
signatures (\emph{transformed}); in \EtwoAStable{}, both fail with the same signature (\emph{stable}). Counting
signature changes raises the overall disagreement from \EtwoADexist{} to \EtwoADall{}.

Different errors do not necessarily mean different repairs. In \IvTrPairs{} transformed pairs, the release-date
installation also stops on the early deprecation. Applying the same handler change to both installations makes
\IvTrBoth{} of these pairs pass under both policies.

The agent checks reinforce this caution. In the fresh samples, agents agree perfectly on first-failure locations (Fleiss'
$\kappa=\ValKappaLoc{}$) but weakly on whether two installations need different migration work
($\kappa=\ValKappaDiff{}$; 95\% CI \ValKappaDiffLo{}--\ValKappaDiffHi{}). In both validation runs, the majority caught
all \ProbeN{} planted errors, although one agent run missed \ProbeAgentMissedMax{}, all of them planted signature
errors. Because the checks lack a human reference standard and signatures do not determine repairs, changed pass/fail
outcomes remain our primary measure.

\subsection{How stable are these findings?}
\label{sec:results-declared}\label{sec:results-manual}

Repeated runs and alternative sampling checks support the main finding. All \EtwoARepeated{} pairs with changed
outcomes or signatures keep their category by majority across three runs. Counting each source commit once gives an
outcome-change rate of \EtwoACommitDexist{}, and vendor-level resampling gives a 95\% interval of
\ClusterThreeLo{}--\ClusterThreeHi{}. Where both external runners work, \EtwoRtwoSame{} of \EtwoRtwoBoth{} pairs
keep the same category under PHPUnit~8.5 and 9.6.

A different runner setup is less directly comparable. Installing PHPUnit through Composer yields an outcome-change
rate of \EtwoBDexist{} in its eligible population (Figure~\ref{fig:e2}, middle row), but also changes other
dependencies in \EtwoMovedBtwo{} release-date and \EtwoMovedAtwo{} modern installations. Restricting to pairs
whose updates stay within runner packages and whose PHPUnit versions match leaves \EtwoPreserving{} pairs:
\EtwoPreservingMasked{} change outcome and two more change signature (Figure~\ref{fig:e2}, bottom row), too few
for a precise rate.

\rqanswer{RQ3}{With the platform check bypassed, PHP~8.3 outcomes differ in \EtwoADexist{} of eligible pairs, nearly
always favoring modern selection, and the two candidate task sets overlap by 58.5\%. Most differences share one
Laravel-related deprecation mechanism; excluding the 130 affected pairs leaves \EtwoMechRestDexist{} disagreement.
Suppressing that deprecation often exposes another failure.}

\section{Implications for Migration Benchmarks}
\label{sec:implications}

A migration benchmark must define what it expects a tool
to fix. A historical benchmark starts with libraries
available when the project was released. A
modernization-first benchmark updates libraries before
migration. Both are valid, but they measure different
tasks. Benchmarks should make this distinction clear
and state whether tools may update dependencies.

For historical reconstruction, passing tests are not
enough. Preserved locks provide the strongest record of
selected versions. Without them, release-date selection
offers a closer approximation (RQ2) and preserves more
passing baselines (RQ1). The Composer version also matters.
Older plugins may require Composer~1. Selection policies,
fallback rules, and unresolved releases should therefore
be documented.

For modernization-first benchmarks, dependency updates
may remove failures before migration begins. Those
failures are no longer part of the evaluated task.
However, passing tests do not guarantee compatibility
with the target runtime. RQ3 also shows that removing
the first failure can expose another. Framework
composition and recurring failure mechanisms help explain
what remains for migration tools to address.

Finally, every benchmark should preserve the source
archive, selected dependency lock, runtime and test-runner
versions, and execution instructions. It should record
platform-check decisions and any dependency changes made
while preparing the test runner. This matters when
harnesses select packages during image
builds~\cite{swebench_php_specs}. Installation failures,
test failures, missing tests, and infrastructure failures
should be reported separately.

A benchmark's results can only be interpreted alongside
its starting environment and the changes its tools
are allowed to make.

\section{Threats to Validity}\label{sec:threats}

\subsection{Construct validity}

A PHP~7.4 CI job indicates intended support, not historical
success. We therefore execute the tests, but passing means
success only under our conditions. Test coverage is unknown,
and missing services may affect outcomes. The RQ1 advantage
persists when all-skipped suites are excluded or usable test
reports are required (Table~\ref{tab:sensitivity}). CI pattern
matching may also miss unusual configurations. We checked
lock classification on 11 constructed and 18 sampled cases,
but did not measure the PHP~7.4 matcher's precision.

A committed lock records selected versions, not necessarily
a working historical environment. We measure version recovery
and test success separately, and examine broader lock groups
and another date cutoff. These checks do not establish when
the original lock was created.

RQ3 measures test behavior, not complete migration readiness.
Bypassing Composer's platform check lets us compare PHP~8.3
test outcomes. We record its verdict separately, but passing
tests do not establish compatibility under the declared
platform constraints. A shared runner reduces runner-related
differences while restricting the eligible population;
alternative-runner checks do not remove this limitation.
Different failure signatures may require the same repair,
and our interventions reveal what follows an initial failure,
not which repairs are necessary.

The primary findings do not depend on LLM judgments. Agent
agreement, checked with fresh samples, repeated coding, and
planted errors, supports some failure interpretations but
cannot establish their correctness without human reference
labels. Agreement is stronger for observable failures than
for judgments about required repairs.

\subsection{Internal validity}

Paired comparisons hold source archives, dependency
requirements, saved metadata, and execution settings fixed.
The primary Composer~2 comparison isolates dependency
selection; Composer~1 is examined separately because older
plugins may require it. Neither setting recreates every
historical toolchain, and recorded release dates may differ
from actual publication times.

Two implementation issues required corrections. Missing
provider and replacer records in the Composer~1 mirror
affected dependency resolution. Correcting them reversed
the secondary Composer-version estimate under release-date
selection, from \OrigToolDiff{} to \ToolDiff{}. We reran both
Composer~1 settings; the primary Composer~2 comparison was
unaffected. Container-name collisions also affected some
RQ3 executions. We fixed the collision and fully reran the
affected procedures. Both corrections and the superseded
results are preserved in the replication package.

Execution uncertainty remains. Package archives may have
changed in availability between runs, and concurrency
exceeded the planned level. Monitoring timeouts, retrying
infrastructure failures, and repeating inconsistent outcomes
reduce, but cannot eliminate, infrastructure-related errors.

Our upgrade experiments reproduce failures after selected
dependency updates, but Composer can also change unrequested
packages. Moreover, similar errors need not have the same
cause. These experiments link permitted upgrades to
observed failures without identifying any individual package
as uniquely responsible.

\subsection{External validity}

The sample requires PHP~7.4 CI configurations and consists
mostly of releases from 2020. Results may differ for older
releases, projects without such CI configurations, or other
package ecosystems. RQ2 uses a separate lock-based sample;
it cannot establish the historical dependencies of RQ1's
releases or represent projects without locks.

RQ3's shared-runner requirement further narrows the eligible
population. Laravel-related projects are overrepresented
among eligible pairs and account for most outcome changes.
The overall disagreement rate may therefore differ in
populations with other framework mixes.

\subsection{Conclusion validity}

Statistical tests treat releases as independent, although
releases may share commits, vendors, or frameworks. Analyses
counting each source commit once and exploratory vendor-level
resampling preserve the main directions, but residual
dependence remains possible.

Framework analyses, comparability checks, and interventions
were added after inspecting RQ3 outcomes and are reported as
exploratory. Study plans were recorded locally rather than
independently preregistered. We use Holm adjustment for RQ2's
secondary comparisons.

\section{Related Work}\label{sec:related}

Our work connects three lines of research: reconstructing old environments, constructing benchmarks, and studying
dependency updates. Each addresses part of our problem, but none measures how dependency selection shapes a
migration benchmark.

\subsection{Historical environment reconstruction}

Rebuilding old software is difficult for Java snapshots
\cite{tufano2017compile,maesbermejo2022revisiting,sulir2016buildability,maesbermejo2024testing}, Python
builds~\cite{mukherjee2021pydfix}, and notebooks~\cite{pimentel2019jupyter}. Inference tools recover missing
requirements from API usage, imports, or dependency knowledge
\cite{wang2021snifferdog,horton2019dockerizeme,ye2022pyego,cheng2024readpye}. Date-based selection also exists:
\texttt{rang} rebuilds R environments at a chosen date~\cite{chan2023rang}, npm studies use archived registry
records~\cite{pinckney2023semver,pinckney2023npmfollower}, functional package managers support temporal
reproducibility~\cite{malka2024nix,malka2025functional}, and CHRONO-RESOLUTION records dependency selections at
release times~\cite{rahman2026chrono}.

This work treats a successful build or test run as the goal. We instead treat dependency selection as the
experimental variable, and we check reconstructed versions against committed locks rather than accepting passing tests
as evidence of historical accuracy (RQ2). Our results also qualify a negative finding: Jin et al.\ report that
date-based backporting does not improve Kaggle notebook reproducibility, using inferred Python versions and
date-constrained imports~\cite{jin2026mlemodernizer}. With PHP held fixed and the complete dependency graph filtered,
release-date selection retains \PrimDiff{} more passing baselines in our setting (RQ1).

\subsection{Benchmark construction}

SWE-bench and its successors build executable environments for issue resolution
\cite{jimenez2024swebench,badertdinov2025swerebench,badertdinov2026swerebenchv2}, and setup agents automate
installation and testing~\cite{eliseeva2025envbench,hu2025repo2run,bouzenia2025executionagent}. MigrationBench and
FreshBrew provide Java migration tasks~\cite{liu2026migrationbench,may2026freshbrew}, and BugsPHP supports PHP
repair~\cite{pramod2024bugsphp}. Some pipelines already restrict versions by date: SWE-bench-Live date-filters its
package proxy~\cite{zhang2025swebenchlive}, and SetUpAgent uses \texttt{uv --exclude-newer}
\cite{vergopoulos2025setupagent,uv_exclude_newer}. Others select versions when building images
\cite{swebench_php_specs}. In all cases, the selection policy is a setup decision; its effect on which tasks the
benchmark contains is not measured.

TimeMachine-bench is the closest work~\cite{fujii2026timemachine}. It builds Python migration tasks with
date-filtered dependencies and reports that 36.8\% of repositories passing in the old environment fail in a new one.
That new environment, however, changes the Python version and unpins dependencies at the same time, so the share due
to dependency selection cannot be separated from the language upgrade. We isolate it. With source and PHP version held
fixed, dependency selection alone changes PHP~7.4 pass rates by \PrimDiff{} (RQ1), and on PHP~8.3 it changes outcomes
in \EtwoADexist{} of eligible pairs, leaving 58.5\% overlap between the two candidate task sets (RQ3).

\subsection{Dependency updates and lock files}

Studies of version constraints and semantic versioning
\cite{decan2021semver,decan2021zero,decan2019ecosystems,dietrich2019versioning}, and of breaking changes in Maven and
npm~\cite{ochoa2022breaking,raemaekers2017semver,jayasuriya2023breaking,venturini2023broke,reyes2024bump}, show that
allowed updates can break clients, and that tests detect only some incompatibilities
\cite{mujahid2020tests,hejderup2022trust}. These studies ask whether an update breaks a client. We ask what that
breakage does to a benchmark: allowed updates remove passing baselines before migration (RQ1) and hide or change
failures after it (RQ3).

Lock-file studies examine how versions are recorded and how past releases can be rebuilt from them
\cite{gamage2026lockfiles,schmid2026mavenlockfile}. Locks cannot settle the question for most PHP releases, however:
a library's Composer lock does not govern consumer installations~\cite{composer_libraries_lock}, and
\LocklessShare{} of releases with manifests in our corpus have no lock at all. For these releases, a benchmark must
choose a selection policy, and our results show that the choice matters.

\textbf{Novelty.} Prior work reconstructs old environments, builds benchmarks under one fixed selection policy, or
studies breaking updates in isolation. To our knowledge, this is the first study to treat dependency selection as the
experimental variable in migration-benchmark construction. We compare selection policies on unchanged source with the
language version held fixed, validate the reconstruction against committed locks, and show that the choice changes
both which projects qualify as baselines and which failures the benchmark exposes.
\section{Conclusion}\label{sec:conclusion}

Our PHP~7.4-to-8.3 study shows that dependency selection
changes more than whether an old project runs.
Release-date selection preserves more passing baselines
and better recovers recorded dependency versions.
Modern selection makes some PHP~8.3 test failures
disappear, particularly in Laravel projects.
Our interventions reveal a deeper effect: removing
the first failure can restore passing tests or expose
another hidden failure.

These findings change how migration benchmarks should
be understood. Historical reconstruction and migration
after dependency modernization represent different
tasks. The chosen starting environment determines
which projects qualify and which problems migration
tools are asked to solve. Dependency selection is
therefore part of benchmark design, not merely setup.

We show that the two policies produce different task
sets, but not yet how much this changes tool scores.
Future work will test whether allowing dependency
updates improves migration-agent success on tasks that
appear under only one policy, and whether tasks shared
by both differ in difficulty across the two dependency
environments.
\section*{Data Availability}

The anonymized replication package will be released upon acceptance. It contains the study inputs and
archive index; the mirror and execution code; the analysis plan with its addenda, deviations and freeze records;
per-project locks, logs, test reports and outcomes; and all label-check materials. The analysis scripts produce the
reported estimates, tables and result figures from the archived outcomes.


\section*{Acknowledgments}

The authors directed the study and reviewed the code and
manuscript. Claude Code (Anthropic, Claude Opus~5.5)
generated and debugged code, ran experiment commands,
and drafted and revised plans and manuscript text.
Section~\ref{sec:design-manual} documents the LLM
label checks. The authors take responsibility for
the methods, results, and conclusions.

\bibliographystyle{IEEEtran}
\bibliography{refs}

@article{newcombe1998paired,
  author  = {Robert G. Newcombe},
  title   = {Improved confidence intervals for the difference between binomial proportions based on paired data},
  journal = {Statistics in Medicine},
  volume  = {17},
  number  = {22},
  pages   = {2635--2650},
  year    = {1998}
}

@article{zeller2002ddmin,
  author  = {Andreas Zeller and Ralf Hildebrandt},
  title   = {Simplifying and isolating failure-inducing input},
  journal = {IEEE Transactions on Software Engineering},
  volume  = {28},
  number  = {2},
  pages   = {183--200},
  year    = {2002},
  doi     = {10.1109/32.988498}
}

@article{tufano2017compile,
  author = {Tufano, Michele and Palomba, Fabio and Bavota, Gabriele and Di Penta, Massimiliano and Oliveto, Rocco and De Lucia, Andrea and Poshyvanyk, Denys},
  title = {{There and back again: Can you compile that snapshot?}},
  journal = {Journal of Software: Evolution and Process},
  volume = {29}, number = {4}, pages = {e1838}, year = {2017},
  publisher = {Wiley}, doi = {10.1002/smr.1838}
}

@article{maesbermejo2022revisiting,
  author = {Maes-Bermejo, Michel and Gallego, Micael and Gortázar, Francisco and Robles, Gregorio and Gonzalez-Barahona, Jesus M.},
  title = {{Revisiting the building of past snapshots — a replication and reproduction study}},
  journal = {Empirical Software Engineering}, volume = {27}, number = {3}, pages = {65}, year = {2022},
  publisher = {Springer}, doi = {10.1007/s10664-022-10117-6}
}

@article{maesbermejo2024testing,
  author = {Maes-Bermejo, Michel and Gallego, Micael and Gortázar, Francisco and Robles, Gregorio and Gonzalez-Barahona, Jesus M.},
  title = {{Testing the past: can we still run tests in past snapshots for Java projects?}},
  journal = {Empirical Software Engineering}, volume = {29}, number = {5}, pages = {127}, year = {2024},
  publisher = {Springer}, doi = {10.1007/s10664-024-10530-z}
}

@inproceedings{sulir2016buildability,
  author = {Sulír, Matúš and Porubän, Jaroslav},
  title = {{A quantitative study of Java software buildability}},
  booktitle = {Proceedings of the 7th International Workshop on Evaluation and Usability of Programming Languages and Tools},
  pages = {17--25}, year = {2016}, publisher = {ACM}, doi = {10.1145/3001878.3001882}
}

@inproceedings{mukherjee2021pydfix,
  author = {Mukherjee, Suchita and Almanza, Abigail and Rubio-González, Cindy},
  title = {{Fixing dependency errors for Python build reproducibility}},
  booktitle = {Proceedings of the 30th ACM SIGSOFT International Symposium on Software Testing and Analysis},
  pages = {439--451}, year = {2021}, publisher = {ACM}, doi = {10.1145/3460319.3464797}
}

@inproceedings{pimentel2019jupyter,
  author = {Pimentel, Joao Felipe and Murta, Leonardo and Braganholo, Vanessa and Freire, Juliana},
  title = {{A Large-Scale Study About Quality and Reproducibility of Jupyter Notebooks}},
  booktitle = {2019 IEEE/ACM 16th International Conference on Mining Software Repositories (MSR)},
  pages = {507--517}, year = {2019}, publisher = {IEEE}, doi = {10.1109/msr.2019.00077}
}

@inproceedings{wang2021snifferdog,
  author = {Wang, Jiawei and Li, Li and Zeller, Andreas},
  title = {{Restoring Execution Environments of Jupyter Notebooks}},
  booktitle = {2021 IEEE/ACM 43rd International Conference on Software Engineering (ICSE)},
  pages = {1622--1633}, year = {2021}, publisher = {IEEE}, doi = {10.1109/icse43902.2021.00144}
}

@inproceedings{horton2019dockerizeme,
  author = {Horton, Eric and Parnin, Chris},
  title = {{DockerizeMe: Automatic Inference of Environment Dependencies for Python Code Snippets}},
  booktitle = {2019 IEEE/ACM 41st International Conference on Software Engineering (ICSE)},
  pages = {328--338}, year = {2019}, publisher = {IEEE}, doi = {10.1109/icse.2019.00047}
}

@inproceedings{ye2022pyego,
  author = {Ye, Hongjie and Chen, Wei and Dou, Wensheng and Wu, Guoquan and Wei, Jun},
  title = {{Knowledge-based environment dependency inference for python programs}},
  booktitle = {Proceedings of the 44th International Conference on Software Engineering},
  pages = {1245--1256}, year = {2022}, publisher = {ACM}, doi = {10.1145/3510003.3510127}
}

@article{cheng2024readpye,
  author = {Cheng, Wei and Hu, Wei and Ma, Xiaoxing},
  title = {{Revisiting Knowledge-Based Inference of Python Runtime Environments: A Realistic and Adaptive Approach}},
  journal = {IEEE Transactions on Software Engineering}, volume = {50}, number = {2}, pages = {258--279}, year = {2024},
  publisher = {IEEE}, doi = {10.1109/tse.2023.3346474}
}

@article{chan2023rang,
  author = {Chan, Chung-hong and Schoch, David},
  title = {{rang: Reconstructing reproducible R computational environments}},
  journal = {PLOS ONE}, volume = {18}, number = {6}, pages = {e0286761}, year = {2023},
  publisher = {Public Library of Science}, doi = {10.1371/journal.pone.0286761}
}

@inproceedings{malka2024nix,
  author = {Malka, Julien and Zacchiroli, Stefano and Zimmermann, Théo},
  title = {{Reproducibility of Build Environments through Space and Time}},
  booktitle = {Proceedings of the 2024 ACM/IEEE 44th International Conference on Software Engineering: New Ideas and Emerging Results},
  pages = {97--101}, year = {2024}, publisher = {ACM}, doi = {10.1145/3639476.3639767}
}

@inproceedings{pinckney2023semver,
  author = {Pinckney, Donald and Cassano, Federico and Guha, Arjun and Bell, Jonathan},
  title = {{A Large Scale Analysis of Semantic Versioning in NPM}},
  booktitle = {2023 IEEE/ACM 20th International Conference on Mining Software Repositories (MSR)},
  pages = {485--497}, year = {2023}, publisher = {IEEE}, doi = {10.1109/msr59073.2023.00073}
}

@inproceedings{pinckney2023npmfollower,
  author = {Pinckney, Donald and Cassano, Federico and Guha, Arjun and Bell, Jonathan},
  title = {{npm-follower: A Complete Dataset Tracking the NPM Ecosystem}},
  booktitle = {Proceedings of the 31st ACM Joint European Software Engineering Conference and Symposium on the Foundations of Software Engineering},
  pages = {2132--2136}, year = {2023}, publisher = {ACM}, doi = {10.1145/3611643.3613094}
}

@article{bouzenia2025executionagent,
  author = {Bouzenia, Islem and Pradel, Michael},
  title = {{You Name It, I Run It: An LLM Agent to Execute Tests of Arbitrary Projects}},
  journal = {Proceedings of the ACM on Software Engineering}, volume = {2}, number = {ISSTA}, pages = {1054--1076}, year = {2025},
  publisher = {ACM}, doi = {10.1145/3728922}
}

@inproceedings{liu2026migrationbench,
  author = {Liu, Linbo and Liu, Xinle and Zhou, Qiang and Chen, Lin and Liu, Yihan and Nguyen, Hoan and Tehrani, Behrooz Omidvar and Shen, Xi and Huan, Jun and Tripp, Omer and Deoras, Anoop},
  title = {{MigrationBench: Repository-Level Code Migration Benchmark from Java 8}},
  booktitle = {Proceedings of the 32nd ACM SIGKDD Conference on Knowledge Discovery and Data Mining V.2},
  pages = {9427--9438}, year = {2026}, publisher = {ACM}, doi = {10.1145/3770855.3817526}
}

@inproceedings{may2026freshbrew,
  author = {May, Victor and Misra, Diganta and Luo, Yanqi and Sridhar, Anjali and Gehring, Justine and {Ribeiro Junior}, Silvio Soares},
  title = {{FreshBrew: A Benchmark for Evaluating AI Agents on Java Code Migration}},
  booktitle = {Proceedings of the 2026 IEEE/ACM 48th International Conference on Software Engineering},
  pages = {944--955}, year = {2026}, publisher = {ACM}, doi = {10.1145/3744916.3773214}
}

@inproceedings{fujii2026timemachine,
  author = {Fujii, Ryo and Morishita, Makoto and Yano, Kazuki and Suzuki, Jun},
  title = {{TimeMachine-bench: A Benchmark for Evaluating Model Capabilities in Repository-Level Migration Tasks}},
  booktitle = {Proceedings of the 19th Conference of the European Chapter of the Association for Computational Linguistics (Volume 1: Long Papers)},
  pages = {8233--8264}, year = {2026}, publisher = {Association for Computational Linguistics}, doi = {10.18653/v1/2026.eacl-long.385}
}

@inproceedings{reyes2024bump,
  author = {Reyes, Frank and Gamage, Yogya and Skoglund, Gabriel and Baudry, Benoit and Monperrus, Martin},
  title = {{BUMP: A Benchmark of Reproducible Breaking Dependency Updates}},
  booktitle = {2024 IEEE International Conference on Software Analysis, Evolution and Reengineering (SANER)},
  pages = {159--170}, year = {2024}, publisher = {IEEE}, doi = {10.1109/saner60148.2024.00024}
}

@inproceedings{pramod2024bugsphp,
  author = {Pramod, Dushan and De Silva, Thushan and Thabrew, Udith and Shariffdeen, Ridwan and Wickramanayake, Sandareka},
  title = {{BugsPHP: A dataset for Automated Program Repair in PHP}},
  booktitle = {Proceedings of the 21st International Conference on Mining Software Repositories},
  pages = {128--132}, year = {2024}, publisher = {ACM}, doi = {10.1145/3643991.3644878}
}

@article{decan2021semver,
  author = {Decan, Alexandre and Mens, Tom},
  title = {{What Do Package Dependencies Tell Us About Semantic Versioning?}},
  journal = {IEEE Transactions on Software Engineering}, volume = {47}, number = {6}, pages = {1226--1240}, year = {2021},
  publisher = {IEEE}, doi = {10.1109/tse.2019.2918315}
}

@article{decan2019ecosystems,
  author = {Decan, Alexandre and Mens, Tom and Grosjean, Philippe},
  title = {{An empirical comparison of dependency network evolution in seven software packaging ecosystems}},
  journal = {Empirical Software Engineering}, volume = {24}, number = {1}, pages = {381--416}, year = {2019},
  publisher = {Springer}, doi = {10.1007/s10664-017-9589-y}
}

@article{ochoa2022breaking,
  author = {Ochoa, Lina and Degueule, Thomas and Falleri, Jean-Rémy and Vinju, Jurgen},
  title = {{Breaking bad? Semantic versioning and impact of breaking changes in Maven Central: An external and differentiated replication study}},
  journal = {Empirical Software Engineering}, volume = {27}, number = {3}, pages = {61}, year = {2022},
  publisher = {Springer}, doi = {10.1007/s10664-021-10052-y}
}

@article{raemaekers2017semver,
  author = {Raemaekers, S. and van Deursen, A. and Visser, J.},
  title = {{Semantic versioning and impact of breaking changes in the Maven repository}},
  journal = {Journal of Systems and Software}, volume = {129}, pages = {140--158}, year = {2017},
  publisher = {Elsevier}, doi = {10.1016/j.jss.2016.04.008}
}

@inproceedings{dietrich2019versioning,
  author = {Dietrich, Jens and Pearce, David and Stringer, Jacob and Tahir, Amjed and Blincoe, Kelly},
  title = {{Dependency Versioning in the Wild}},
  booktitle = {2019 IEEE/ACM 16th International Conference on Mining Software Repositories (MSR)},
  pages = {349--359}, year = {2019}, publisher = {IEEE}, doi = {10.1109/msr.2019.00061}
}

@article{decan2021zero,
  author = {Decan, Alexandre and Mens, Tom},
  title = {{Lost in zero space – An empirical comparison of 0.y.z releases in software package distributions}},
  journal = {Science of Computer Programming}, volume = {208}, pages = {102656}, year = {2021},
  publisher = {Elsevier}, doi = {10.1016/j.scico.2021.102656}
}

@article{venturini2023broke,
  author = {Venturini, Daniel and Cogo, Filipe Roseiro and Polato, Ivanilton and Gerosa, Marco A. and Wiese, Igor Scaliante},
  title = {{I Depended on You and You Broke Me: An Empirical Study of Manifesting Breaking Changes in Client Packages}},
  journal = {ACM Transactions on Software Engineering and Methodology}, volume = {32}, number = {4}, pages = {1--26}, year = {2023},
  publisher = {ACM}, doi = {10.1145/3576037}
}

@inproceedings{jayasuriya2023breaking,
  author = {Jayasuriya, Dhanushka and Terragni, Valerio and Dietrich, Jens and Ou, Samuel and Blincoe, Kelly},
  title = {{Understanding Breaking Changes in the Wild}},
  booktitle = {Proceedings of the 32nd ACM SIGSOFT International Symposium on Software Testing and Analysis},
  pages = {1433--1444}, year = {2023}, publisher = {ACM}, doi = {10.1145/3597926.3598147}
}

@inproceedings{mujahid2020tests,
  author = {Mujahid, Suhaib and Abdalkareem, Rabe and Shihab, Emad and McIntosh, Shane},
  title = {{Using Others' Tests to Identify Breaking Updates}},
  booktitle = {Proceedings of the 17th International Conference on Mining Software Repositories},
  pages = {466--476}, year = {2020}, publisher = {ACM}, doi = {10.1145/3379597.3387476}
}

@article{hejderup2022trust,
  author = {Hejderup, Joseph and Gousios, Georgios},
  title = {{Can we trust tests to automate dependency updates? A case study of Java Projects}},
  journal = {Journal of Systems and Software}, volume = {183}, pages = {111097}, year = {2022},
  publisher = {Elsevier}, doi = {10.1016/j.jss.2021.111097}
}

@article{gamage2026lockfiles,
  author = {Gamage, Yogya and Tiwari, Deepika and Monperrus, Martin and Baudry, Benoit},
  title = {{The design space of lockfiles across package managers}},
  journal = {Empirical Software Engineering}, volume = {31}, number = {3}, pages = {63}, year = {2026},
  publisher = {Springer}, doi = {10.1007/s10664-025-10789-w}
}

@inproceedings{schmid2026mavenlockfile,
  author = {Schmid, Larissa and Lundell, Elias and Gamage, Yogya and Baudry, Benoit and Monperrus, Martin},
  title = {{Maven-Lockfile: High Integrity Rebuild of Past Java Releases}},
  booktitle = {Proceedings of the 2026 IEEE/ACM 48th International Conference on Software Engineering},
  series = {ICSE-Companion '26},
  pages = {97--100}, year = {2026}, publisher = {ACM}, doi = {10.1145/3774748.3787615}
}

@inproceedings{malka2025functional,
  author = {Malka, Julien and Zacchiroli, Stefano and Zimmermann, Théo},
  title = {{Does Functional Package Management Enable Reproducible Builds at Scale? Yes.}},
  booktitle = {2025 IEEE/ACM 22nd International Conference on Mining Software Repositories (MSR)},
  pages = {775--787}, year = {2025}, publisher = {IEEE}, doi = {10.1109/msr66628.2025.00115}
}

@inproceedings{jimenez2024swebench,
  author = {Jimenez, Carlos E. and Yang, John and Wettig, Alexander and Yao, Shunyu and Pei, Kexin and Press, Ofir and Narasimhan, Karthik},
  title = {{SWE-bench: Can Language Models Resolve Real-World GitHub Issues?}},
  booktitle = {International Conference on Learning Representations (ICLR 2024)},
  year = {2024}, url = {https://openreview.net/forum?id=VTF8yNQM66}, eprint = {2310.06770}, archivePrefix = {arXiv}
}

@inproceedings{badertdinov2025swerebench,
  author = {Badertdinov, Ibragim and Golubev, Alexander and Nekrashevich, Maksim and Shevtsov, Anton and Karasik, Simon and Andriushchenko, Andrei and Trofimova, Maria and Litvintseva, Daria and Yangel, Boris},
  title = {{SWE-rebench: An Automated Pipeline for Task Collection and Decontaminated Evaluation of Software Engineering Agents}},
  booktitle = {Advances in Neural Information Processing Systems 38 (NeurIPS 2025), Datasets and Benchmarks Track},
  year = {2025}, eprint = {2505.20411}, archivePrefix = {arXiv}
}

@inproceedings{badertdinov2026swerebenchv2,
  author = {Badertdinov, Ibragim and Nekrashevich, Maksim and Shevtsov, Anton and Golubev, Alexander},
  title = {{SWE-rebench V2: Language-Agnostic SWE Task Collection at Scale}},
  booktitle = {Proceedings of the 43rd International Conference on Machine Learning},
  series = {Proceedings of Machine Learning Research},
  volume = {306}, pages = {4820--4843}, publisher = {PMLR},
  year = {2026}, eprint = {2602.23866}, archivePrefix = {arXiv},
  url = {https://proceedings.mlr.press/v306/badertdinov26a.html}
}

@inproceedings{zhang2025swebenchlive,
  author = {Zhang, Linghao and He, Shilin and Zhang, Chaoyun and Kang, Yu and Li, Bowen and Xie, Chengxing and Wang, Junhao and Wang, Maoquan and Huang, Yufan and Fu, Shengyu and Nallipogu, Elsie and Lin, Qingwei and Dang, Yingnong and Rajmohan, Saravan and Zhang, Dongmei},
  title = {{SWE-bench Goes Live!}},
  booktitle = {Advances in Neural Information Processing Systems 38 (NeurIPS 2025), Datasets and Benchmarks Track},
  year = {2025}, eprint = {2505.23419}, archivePrefix = {arXiv}
}

@inproceedings{vergopoulos2025setupagent,
  author = {Vergopoulos, Konstantinos and M{\"u}ller, Mark Niklas and Vechev, Martin},
  title = {{Automated Benchmark Generation for Repository-Level Coding Tasks}},
  booktitle = {Proceedings of the 42nd International Conference on Machine Learning (ICML)},
  series = {Proceedings of Machine Learning Research}, volume = {267}, pages = {61240--61265}, year = {2025},
  url = {https://proceedings.mlr.press/v267/vergopoulos25a.html}
}

@misc{eliseeva2025envbench,
  author = {Eliseeva, Aleksandra and Kovrigin, Alexander and Kholkin, Ilia and Bogomolov, Egor and Zharov, Yaroslav},
  title = {{EnvBench: A Benchmark for Automated Environment Setup}},
  year = {2025}, eprint = {2503.14443}, archivePrefix = {arXiv}, primaryClass = {cs.LG},
  note = {DL4Code workshop at ICLR 2025}
}

@inproceedings{hu2025repo2run,
  author = {Hu, Ruida and Peng, Chao and Wang, Xinchen and Xu, Junjielong and Gao, Cuiyun},
  title = {{Repo2Run: Automated Building Executable Environment for Code Repository at Scale}},
  booktitle = {Advances in Neural Information Processing Systems 38 (NeurIPS 2025)},
  year = {2025}, eprint = {2502.13681}, archivePrefix = {arXiv}
}

@misc{rahman2026chrono,
  author = {Rahman, Imranur and Marley, Jill and Paramitha, Ranindya and Williams, Laurie},
  title = {{CHRONO-RESOLUTION: A Dependency Resolution Dataset at Release Points for npm, PyPI, and crates.io Packages}},
  year = {2026}, eprint = {2607.15315}, archivePrefix = {arXiv}, primaryClass = {cs.SE},
  note = {Accepted to ICSME 2026, Tool Demonstration and Data Showcase Track}
}

@article{jin2026mlemodernizer,
  author = {Jin, Bihui and Wang, Kaiyuan and Nie, Pengyu},
  title = {{Automated Modernization of Machine Learning Engineering Notebooks for Reproducibility}},
  journal = {Proceedings of the ACM on Software Engineering}, volume = {3}, number = {ISSTA}, pages = {1718--1739}, year = {2026},
  doi = {10.1145/3832166}, eprint = {2602.07195}, archivePrefix = {arXiv}
}

@misc{swebench_php_specs,
  author = {{SWE-bench maintainers}},
  title = {{SWE-bench harness: PHP environment specifications (tag v4.0.4)}},
  howpublished = {\url{https://github.com/SWE-bench/SWE-bench/blob/v4.0.4/swebench/harness/constants/php.py}},
  note = {Accessed 2026-10-01}
}

@misc{composer_libraries_lock,
  author = {{Composer}},
  title = {{Libraries: lock file}},
  howpublished = {\url{https://getcomposer.org/doc/02-libraries.md\#lock-file}},
  note = {Accessed 2026-10-01}
}

@misc{uv_exclude_newer,
  author = {{Astral}},
  title = {{uv documentation: reproducible resolutions (exclude-newer)}},
  howpublished = {\url{https://docs.astral.sh/uv/concepts/resolution/\#reproducible-resolutions}},
  note = {Accessed 2026-10-01}
}

@inproceedings{ahmed2025llm,
  author = {Ahmed, Toufique and Devanbu, Premkumar and Treude, Christoph and Pradel, Michael},
  title = {{Can LLMs Replace Manual Annotation of Software Engineering Artifacts?}},
  booktitle = {2025 IEEE/ACM 22nd International Conference on Mining Software Repositories (MSR)},
  pages = {526--538}, year = {2025}, publisher = {IEEE}, doi = {10.1109/MSR66628.2025.00086}
}

@misc{composer_commands,
  author       = {{Composer}},
  title        = {{Command-line interface / Commands}},
  howpublished = {\url{https://getcomposer.org/doc/03-cli.md}},
  note         = {Accessed 2026-10-08}
}

@misc{composer_plugins,
  author       = {{Composer}},
  title        = {{Setting up and using plugins}},
  howpublished = {\url{https://getcomposer.org/doc/articles/plugins.md}},
  note         = {Accessed 2026-10-10}
}

@misc{composer_upgrade2,
  author       = {{Composer}},
  title        = {{Upgrading from Composer 1 to 2 (UPGRADE-2.0.md)}},
  howpublished = {\url{https://github.com/composer/composer/blob/main/UPGRADE-2.0.md}},
  note         = {Accessed 2026-10-10}
}

@misc{composer_minifier,
  author       = {{Composer}},
  title        = {{composer/metadata-minifier}},
  howpublished = {\url{https://github.com/composer/metadata-minifier}},
  note         = {Accessed 2026-10-10}
}

@misc{composer_platform,
  author       = {{Composer}},
  title        = {{Composer platform dependencies}},
  howpublished = {\url{https://getcomposer.org/doc/articles/composer-platform-dependencies.md}},
  note         = {Accessed 2026-10-08}
}

@misc{php80_incompatible,
  author       = {{The PHP Documentation Group}},
  title        = {{Migrating from PHP 7.4.x to PHP 8.0.x:
                   Backward Incompatible Changes}},
  howpublished = {\url{https://www.php.net/manual/en/migration80.incompatible.php}},
  note         = {Accessed 2026-10-08}
}

@misc{php81_incompatible,
  author       = {{The PHP Documentation Group}},
  title        = {{Migrating from PHP 8.0.x to PHP 8.1.x:
                   Backward Incompatible Changes}},
  howpublished = {\url{https://www.php.net/manual/en/migration81.incompatible.php}},
  note         = {Accessed 2026-10-08}
}

@misc{composer_repositories,
  author       = {{Composer}},
  title        = {{Repositories}},
  howpublished = {\url{https://getcomposer.org/doc/05-repositories.md}},
  note         = {Accessed 2026-10-08}
}

\end{document}